\documentclass[preprint,endfloats,pre,draft]{revtex4}
\usepackage{bm}
\usepackage[final]{graphicx}
\usepackage{longtable}
\newcommand{\be}{\begin{equation}}
\newcommand{\ee}{\end{equation}}
\newcommand{\bea}{\begin{eqnarray}}
\newcommand{\eea}{\end{eqnarray}}

\begin{document}

\title{Influence of long-range correlated quenched disorder on the 
adsorption of long flexible polymer chains on a wall.}

\author{Z. Usatenko$^{1,2}$, A.Ciach$^2$}

\affiliation{$^{1}$Institute for Condensed Matter Physics,
National Academy of Sciences of Ukraine, 1~Svientsitskii Str.,
UA--79011 Lviv, Ukraine }
\email{pylyp@ph.icmp.lviv.ua} \affiliation{$^{2}$Institute of
Physical Chemistry of the Polish Academy of Sciences, Warsaw,
01-224, Poland}



\begin{abstract}

The process of adsorption on a planar wall of long-flexible
polymer chains in the medium with quenched long-range correlated
disorder is investigated. We focus  on the case of correlations
between defects or impurities that decay according to the
power-low $ x^{-a}$ for large distances $x$, where ${\bf x}=({\bf
r},z)$.  Field theoretical approach in $d=4-\epsilon$ and directly
in $d=3$ dimensions up to one-loop order for the semi-infinite
$|\phi|^4$ m-vector model (in the limit $m\to 0$) with a planar
boundary is used. The whole set of surface critical exponents at
the adsorption threshold $T=T_a$, which separates the nonadsorbed
region from the adsorbed one is obtained. Moreover, we calculate
the crossover critical exponent $\Phi$ and the set of exponents
associated with them. We perform calculations in a double
$\epsilon=4-d$ and $\delta=4-a$ expansion and also for a fixed
dimension $d=3$, up to one-loop order for different values of the
correlation parameter $2<a\le 3$.

 The obtained results indicate that for the systems with long-range
 correlated quenched disorder the new set of surface critical exponents arises.
All the surface critical exponents depend on $a$. Hence,
 the presence of long-range correlated disorder influences the process of adsorption of
 long-flexible polymer chains on a wall in a significant way.

\end{abstract}

\maketitle
\section{Introduction}
   Universal properties of long-flexible polymer chains change when
   small amount of long-range correlated quenched disorder is
   introduced into an infinite medium
   \cite{Blavatska01,Holovatch01}. The correlated defects
   (i.e. regions that cannot be occupied by the chain) may occur, for
   example, in a porous medium or in a disordered sponge-like
   structure formed by lipid membranes in biological
   systems. Intuitively, if the correlations between the defects
   and/or impurities decay sufficiently slowly, then the chain has to
   go around large correlated regions, and effectively occupies larger
   space, with the defects contained inside the region occupied by the
   coil. As a result, the polymer swells. If, however, the range of
   correlations is very large, then the polymer may be trapped between
   the walls of defects (i.e. the probability of going beyond the
   defected region is low), and this may lead to a collapsed
   state. These heuristic arguments suggest that the polymer either
   swells or collapses, depending on the range of correlations between
   the defects or impurities. Indeed, recent results agree with the
   intuitive expectations \cite{Blavatska01,Holovatch01}.  For
   different ranges of correlations (i.e.  different values of $a$ for
   the power-law decay of correlations $1/x^a$) the swelling of the
   polymer is described by different dependencies of the radius of
   gyration on the number of monomers. Finally, for $a < 2.3$
   \cite{Blavatska01,Holovatch01} a first-order transition to a
   collapsed state was found \cite{Blavatska01,Holovatch01}.

Motivated by the above results we focus our attention on the effect of
the presence of a small amount of long-range correlated quenched
disorder in the bulk on the adsorption of long-flexible polymer chains
on a planar surface forming the system boundary.
 In real systems different kinds of
 defects and impurities may be localized inside the bulk or at the
 boundary.  As was found in Ref.\cite{DN89}, introducing into the
 system {\it short-range correlated random quenched surface disorder}
 is irrelevant for critical behavior, but long-range correlated
 quenched surface disorder with $ g(r)\sim 1/r^{a} $ can be relevant
 if $a<d-1$, and is irrelevant if $a\geq d-1$.  The question how the
 adsorption phenomena of long flexible polymer chains depend on the
 presence of long-range-correlated quenched disorder in the {\it bulk}
 remains open, however.  New universality class characterizing the
 polymer in the presence of long-range correlated disorder indicates
 that the critical exponents describing the properties of the polymer
 chain near the wall should assume different values than in the pure
 system. The purpose of this work is a determination of the surface
 critical exponents to first order in perturbation expansion, in order
 to gain information about a qualitative change of adsorption of the
 chains when the range of correlations between the defects in the bulk
 increases (i.e., $a$ is decreased).

Long flexible polymer chains in a good solvent are perfectly
described by a model of self-avoiding walks (SAW) on a regular
lattice \cite{Cloizeaux}. Their scaling properties in the limit of
an infinite number of steps $N$ may be derived by a formal $m \to 0$
limit of the $O(m)$ vector model at its critical point
\cite{deGennes}.  The average square
end-to-end distance, the number of configurations with one end fixed
and with both ends fixed at the distance $x=\sqrt{({\vec x}_{A}-{\vec
x}_{B})^{2}}$ exhibit the following asymptotic behavior in the limit
$N\to \infty$ \be <R^2>\sim N^{2\nu},\quad\quad\quad Z_{N}\sim
q^{N}N^{\gamma-1},\quad\quad\quad Z_{N}(x)\sim
q^{N}N^{-(2-\alpha)},\label{RZ}
\ee
 respectively.  $\nu$, $\gamma$ and $\alpha$ are the universal
 correlation length, susceptibility and specific heat critical
 exponents for the $m=0$ model, $d$ is the space dimensionality, $q$
 is a non universal fugacity. $1/N$ plays a role of a critical
 parameter  analogous to the reduced  critical temperature in
 magnetic systems.

When the polymer solution is in contact with a
 solid substrate (or with vapor), then the monomers interact with the
 surface (or their chemical potential at the interface is different
 than in the bulk). At sufficiently low temperatures, $T<T_a$, the
 attraction between the monomers and the surface leads to an adsorbed
 state, where a finite fraction of the monomers is attached to the
 system boundary. Deviation from the adsorption threshold,
 $c\propto(T-T_a)/T_a)$, changes sign at the transition between the adsorbed
 ($c<0$) and the nonadsorbed state ($c>0$) and it plays a role of a
 second critical parameter. The adsorption threshold for infinite
 chains, where $1/N\to 0$ and $c\to 0$ is a multicritical phenomenon.
 We shall assume that the solution of polymer chains is sufficiently
 dilute, so that interchain interactions and overlapping between
 different chains can be neglected, and it is sufficient to consider
 surface effects for configurations of a single chain. For pure
 solvents the investigation of adsorption phenomena of long-flexible
 polymer chains on a surface was a subject of a series of works (for
 the sake of brevity we notice only few of them
\cite{deGennes,EKB82,eisenriegler:83,Eisenriegler,HG94,SKG99,SGK01,RDGKS02,ZLB90}).
 The polymer adsorption on a wall in the
limit of an infinite chain is closely related to surface critical
phenomena in the $m$ - vector model of a magnet in a semi-infinite
geometry in the limit $m
\to 0$ \cite{G76,deGennes,Barber}. Based on the above analogy,
Eisenriegler and co-workers
\cite{EKB82,eisenriegler:83,Eisenriegler} described the scaling properties
 of long chains
near a wall on the basis of the results of the field theory
developed for semi-infinite magnetic systems in
Ref.\cite{DD81,DD83,D86}. Surface multicritical phenomena in
dilute polymer systems (at $T= T_a$ and $N\to \infty$) correspond
to the special transition in semiinfinte magnets. The special
transition ($c=0$) is characterized by one additional independent
surface critical exponent $\eta_{\parallel}$, which characterizes
critical correlations in directions parallel to the surface. The
whole set of the other surface critical exponents can be obtained
on the basis of $\eta_{\parallel}$ and the bulk critical exponents
$\nu$ and $\eta$ with the help of surface scaling relations. The
crossover critical exponent $\Phi$ characterizes the crossover
behavior between the special and ordinary transitions ($c\ne 0$).
The latter exponent is related to the length scale
\cite{eisenriegler:83,Eisenriegler}, \be \label{xic} \xi_c \sim
|c|^{-\nu/\Phi}, \ee
 associated with the parameter $c$.  In the polymer problem the length
 $\xi_c$ can be interpreted as the distance from the surface up to
 which the properties of the polymer depend on the value of $c$, not
 only on its sign.  The remaining, bulk length scales are the average
 end-to-end distance $\xi_R=\sqrt{<R^2>}\sim N^{\nu}$ and the
 microscopic length $l$ -- the effective monomer linear
 dimension. Near the multicritical point the only relevant lengths are
 $\xi_R\to\infty$ and $\xi_c\to\infty$, and the properties of the
 system depend on the ratio $\xi_R/\xi_c$. In the asymptotic scaling
 regime the universal physical quantities $X(N,c)$ and $Y(z;N,c)$
 assume the scaling forms
\be
\label{scal}
X(N,c)=N^{a_X}X^s_{\pm}\big(\xi_R/\xi_c\big),\hskip1cm
Y(z;N,c)=N^{a_Y}Y^s_{\pm}\big(z/\xi_R,\xi_R/\xi_c\big),
\ee
where $X^s_{\pm}$ and $Y^s_{\pm}$ denote the scaling functions with
the subscripts $+$ and $-$ corresponding to $c>0$ and $c<0$
respectively.  The characteristic length ratio is
$(\xi_R/\xi_c)^{\Phi/\nu} \sim |c|N^{\Phi}$, where $cN^{\Phi}$ is
the standard scaling variable
\cite{EKB82}.  The exponents $a_X$ and $a_Y$  assume
 different values for different quantities $X$ and $Y$. Let us first
 consider the mean square end-to-end distance for one end attached to
 the surface and the other one free. In the semi-infinite system the
 translational invariance is broken, and the parallel $
 <R^2_{\parallel}>$ and perpendicular $<R^2_{\perp}>$ parts of the
 average end-to-end distance $<R^2>=<R^2_{\perp}+R^2_{\parallel}>$
 should be distinguished. For $<R^2_{\perp}>^{1/2}$ the exponent in
 the scaling form (\ref{scal}) is $a_X=\nu$ and  the corresponding scaling
 functions assume the form $\sim const$ for $c\ge 0$ and
 $\sim 1/y$
 for $c< 0$, where $y=\xi_R/\xi_c$ \cite{EKB82}.  Thus, for the
 adsorbed state and for $N\to\infty$ the length associated with $c$
 describes the thickness $\xi$ of the adsorbed layer,

\be
\label{xi}
\xi = <R^2_{\perp}>^{1/2}\sim\xi_c \hskip1cm c<0.
\ee
This thickness diverges for $c=1/N=0$ and for finite negative values
of $c$ remains finite for an infinite chain. For $c\ge 0$ the
asymptotic behavior of the mean distance of the free end from the
other end attached to the surface is
\be
\label{xi>0}
 <R^2_{\perp}>^{1/2}\sim N^{\nu}\hskip1cm c \ge 0,
\ee
 i.e it has the same asymptotic behavior as in the bulk. The
 asymptotic scaling form of $ <R^2_{\parallel}>^{1/2}$ for $c<0$ is
 $<R^2_{\parallel}>^{1/2}\sim
 |c|^{(\nu^{d-1}-\nu)/\Phi}N^{\nu^{d-1}}$, where $\nu^{d-1}$ is the
 correlation exponent in $d-1$ dimensions. For $c\ge 0$ the scaling
 form of $<R^2_{\parallel}>^{1/2}$ is given by Eq.  (\ref{xi>0}),
 i.e. it is also the same as in the bulk.

For the fraction of monomers at the surface, $N_1/N$, the following
asymptotic behavior has been found for $N\to\infty$
\cite{EKB82,Eisenriegler},
 \begin{eqnarray}
\label{N1/N}
N_{1}/N\sim \left\{
 \begin{array}{lll}
|c|^{(1-\Phi)/\Phi} &\;\; \mbox{if}  & \;\;\; \mbox{$c<0$}\\
N^{\Phi-1} &\;\; \mbox{if}& \;\;\;\mbox{$c=0$}\\
(cN)^{-1} &\;\; \mbox{if}& \;\;\;\mbox{$c>0$}
 \end{array}
 \right. .
 \end{eqnarray}
 Hence, for $N\to\infty$ and for finite, negative values of $c$,
 $N_{1}/N$ is finite, but for $c\ge 0$ $N_{1}/N\to 0$ for
 $N\to\infty$.
The thickness of the adsorbed layer is closely related to the
 fraction of monomers at the surface $N_1/N$ \cite{EKB82,Eisenriegler}, since
 the more monomers are fixed at the wall, the smaller the region
 occupied by the remaining monomers. In particular, for weakly adsorbed
 phase ($ c < 0$ and $|c|\ll 1$) we find $N_{1}/N\sim\xi^{-(1-\Phi)/\nu}$.

 The scaling behavior is also obeyed by the mean number of the
free ends in the layer between $z$ and $z+dz$, which is
proportional to the partition function of a chain with one end
fixed at ${\bf x}_{A}=({\bf r}_{A},z)$ and the other end free,
$Z_{N}(z)$. The density of monomers in a layer  at the distance
$z$ from the wall to which one end of the polymer is attached,
$M_{N}^{\lambda}(z)$
 scales according to Eq. (\ref{scal}) as well.  For
the above quantities the exponent $a_Y$ in (\ref{scal}) is $\gamma-1$
and $\gamma_1-\nu$ respectively. Short-distance behavior ($l\ll z\ll
\xi_R$) of the two quantities right at the threshold ($c=0$) is
\be
\label{ZN}
Z_{N}(z) \sim
z^{(\gamma-\gamma_{1})/ \nu} N^{\gamma_{1}-1}
\ee
 and
\be
\label{MN}
M_{N}^{\lambda}(z)\sim z^{-1+(1-\Phi)/\nu}N^{\gamma_{1} -1+\Phi}.
\ee
  $\gamma_1$
as well as the whole set of the surface critical exponents can be
obtained from $\eta_{\parallel}$ and $\Phi$ through scaling relations
(see the Appendix).  The remaining quantities characterizing the
adsorption process are described in detail in
Ref.\cite{EKB82,Eisenriegler}.

Taking into account the results of ref.
\cite{Blavatska01,Holovatch01} we conclude that for $c\ge 0$ the
polymer with one end attached to the surface swells as in the bulk
when $a$ decreases (see (\ref{xi>0})). However, in order to
determine the effect of the long-range correlated disorder on the
adsorption of the polymer right at the threshold (see (\ref{ZN})
and (\ref{MN})) or in the crossover region (see (\ref{xic}),
(\ref{xi})  and (\ref{N1/N})) it is necessary to find the
dependence of the surface critical exponents on $a$.

In the next section the model is briefly described. In sec.III the
surface multicritical behavior of the system with long-range
correlated disorder  is outlined. The results of sec. III enable
us to obtain in sec. IV the surface critical exponents to first
order in the perturbation expansion. Final section contains a
brief discussion of the results.

\section{the model}
When a disorder is introduced into an infinite magnetic system,
the Landau-Ginzburg-Wilson Hamiltonian \cite{deGennes} assumes the
form

\be H = \int_{V} d^{d}x [\frac{1}{2} \mid \nabla\vec{\phi} \mid
^{2} + \frac{1}{2} (\mu_{0}^{2}+\delta\tau({\bf{x}}))\mid
\vec{\phi} \mid^{2} +\frac{1}{4!} v_{0}
(\vec{\phi}^2)^{2}],\label{1} \ee

where $\vec{\phi}(x)$ is an $m$-vector field with the components
$\phi_{i}(x)$, $i=1,...,m$. Here $\mu_{0}^2$ is the "bare mass",
which in the case of a magnet corresponds to the reduced
temperature. The inhomogeneities in the system cause local
deviations from the average value of the transition temperature,
and $\delta\tau({\bf{x}})$ represents the quenched
random-temperature disorder, with $<\delta\tau({\bf{x}})>=0$ and
\be
\frac{1}{8}<\delta\tau({\bf{x}})\delta\tau({\bf{x}}')>=g(\mid
{\bf{x}}\mid), \label{2}
\ee
 where angular brackets $<...>$ denote
configurational averaging over quenched disorder. Following
Refs.\cite{Blavatska01,Holovatch01} we assume that the pair
correlation function $g(\mid {\bf{x}}\mid)$ falls off with the
distance as \be g(\mid {\bf{x}}\mid) \sim \frac{1}{x^a}\label{2b}
\ee
for large ${\bf{x}}=({\bf{r}},z)$, where $a$ is a constant and
$x=\mid \bf{x}\mid$.
The Fourier-transform ${\tilde{g}}(k)$ of $g(x)$ for small $k$ is
\be
{\tilde g}(k) \sim u_{0}+w_{0} \mid k\mid^{a-d}.\label{3}
\ee
This corresponds to the so-called long-range-correlated
"random-temperature" disorder. In the case of random uncorrelated
point-like (or short-range-correlated) disorder the site-occupation
correlation function is $g(x) \sim \delta(x)$ and its
Fourier-transform  assumes the simple form \be {\tilde g}(k) \sim
u_{0}.\label{4} \ee

Applying the replica method in order to average the free energy
over different configurations of the quenched disorder, it is
possible to construct an effective Hamiltonian of the $|\phi|^4$
$m$ -vector model with a long-range-correlated disorder
\bea
H_{eff} & = & \sum_{\alpha=1}^{n}\int_{V} d^{d}x [\frac{1}{2} \mid
\nabla\vec{\phi}_{\alpha} \mid ^{2} + \frac{1}{2}
\mu_{0}^{2}\vec{\phi}_{\alpha}^{2} +\frac{1}{4!} v_{0}
(\vec{\phi}_{\alpha}^2)^{2}\nonumber\\
&& -\sum_{\alpha,\beta=1}^{n}\int
d^{d}x_{1} d^{d}x_{1}^{'}g(\mid
x_{1}-x_{1}^{'}\mid)\vec{\phi}_{\alpha}^{2}
(x_{1})\vec{\phi}_{\beta}^{2}(x_{1}^{'}).\label{5}
\eea
Here Greek indices denote replicas, and the replica limit $n\to 0$
is implied. If $a\geq d$, then the $w_{0}$ term is irrelevant.
This corresponds to {\it random uncorrelated point-like disorder}
(or short-range-correlated random disorder). As noticed by Kim
\cite{Kim}, in this case in the limit $m,n \to 0$ both $v_{0}$ and
$u_{0}$ terms are of the same symmetry. It indicates that {\it a
weak quenched uncorrelated disorder is irrelevant for SAWs}
\cite{Harris}. If, on the other hand, $a<d$, the term
$w_{0}k^{a-d}$ is relevant for the critical behavior at $k\to 0$,
and {\it the long-range-correlated disorder is relevant for SAWs}
(see \cite{Blavatska01,Holovatch01}).
The limit $m\to 0$ of this model can be interpreted as a model of
long-flexible polymer chains in a disordered medium
\cite{Blavatska01,Holovatch01}.

The presence of a hard wall leads to a modification of the
interactions in the near-surface layer. Thus, in the semi-infinite
system there should be additional, surface contribution to the
Hamiltonian. The effective Hamiltonian of the semi-infinite
$|\phi|^4$ $m$-vector model with long-range-correlated disorder in
this case is

\bea H_{eff} & = & \sum_{\alpha=1}^{n}\int_{V} d^{d}x [\frac{1}{2}
\mid \nabla\vec{\phi}_{\alpha} \mid ^{2} + \frac{1}{2}
\mu_{0}^{2}\vec{\phi}_{\alpha}^{2} +\frac{1}{4!} v_{0}
(\vec{\phi}_{\alpha}^2)^{2}\nonumber\\
&& -\sum_{\alpha,\beta=1}^{n}\int d^{d}x_{1}
d^{d}x_{1}^{'}\bar{g}(\mid
x_{1}-x_{1}^{'}\mid)\vec{\phi}_{\alpha}^{2}
(x_{1})\vec{\phi}_{\beta}^{2}(x_{1}^{'})+\frac{c_{0}}{2}\sum_{\alpha=1}^{n}
\int_{\partial V}d^{d-1}r
\vec{\phi}_{\alpha}^{2}({\bf{r}},z=0),\label{9} \eea
where $c_{0}$ describes the surface-enhancement of interactions.
In the polymer analog $c_0\propto (T-T_a)/T_a$, as already noted in the
introduction. The surface introduces an anisotropy into the problem,
and directions parallel and perpendicular to the surface are no longer
equivalent. In accordance with the fact that we have to deal with
semi-infinite geometry $({\bf x}=({\bf r},z\geq 0))$, only parallel
Fourier transforms in $d-1$ dimensions will be performed. The parallel
Fourier transform ${\tilde{g}}(q,z)$ of (\ref{2}) is
\be {\tilde{g}}(q,z)=
w_{0}\frac{2^{\frac{a-d+1}{2}}}{\Gamma[\frac{d-a}{2}]\sqrt{\pi}}
q^{\frac{a-d+1}{2}}z^{\frac{d-a-1}{2}} K_{\frac{a-d+1}{2}}(q
z),\label{7} \ee
where $K_{\frac{a-d+1}{2}}(q z)$ is the modified Bessel function
and $q=\mid{\bf q}\mid$, where ${\bf q}$ is a $d-1$ dimensional
vector. In the case of small $q$ and $z$ we obtain the relation
\be {\tilde g}(q,z)\sim
u_{0}+w_{0}q^{a-d+1}+w_{0}^{'}z^{d-a-1},\label{8} \ee
which agrees with the predictions obtained in \cite{DN89}. We
concentrate our attention on the case $a\le 3$ for $d=3$, for
which the long-range correlated disorder in the bulk is relevant.
In the general case of arbitrary $z$ (from $z=0$ on the wall to
$z\to \infty$) we must take into account the Fourier transform
${\tilde{g}}(q,z)$ of the form (\ref{7}).
\section{Surface critical behavior near the multicritical point $c=c_a$}
\subsection{Normalization conditions}
The correlation function which involves $N$ fields
$\phi({\bf{x}}_{i})$ at distinct points ${\bf{x}}_{i}(1\leq i \leq
N)$ in the bulk, $M$ fields $\phi({\bf{r}}_{j},z=0)\equiv
\phi_{s}({\bf{r}}_{j})$ at distinct points on the wall with
parallel coordinates ${\bf{r}}_{j}(1\leq j \leq M)$, and $L_{1}$
insertions of the surface operator
$\frac{1}{2}\phi_{s}^{2}({\bf{R}}_{l})$ at points ${\bf{R}}_{l}$
with $1\leq l \leq L_{1}$, has the form

\be G^{(N,M,L_1)}(\{{\bf x}_{i}\},\{{\bf r}_j\},\{{\bf{R}}_{l}\})
= < \prod_{i=1}^{N} \phi({\bf x}_{i})\prod_{j=1}^{M}\phi_{s}({\bf
r}_{j})\prod_{l=1}^{L_{1}}\frac{1}{2}\phi^{2}_{s}({\bf{R}}_{l})>,
\label{10} \ee
where $<...>$ denotes averaging with the Boltzmann factor, in
which the Hamiltonian is given in Eq.(\ref{9}). The corresponding
full free propagator in the mixed ${\bf{p}} z$ representation is
given by \cite{D86}

\be G({{\bf{p}}},z,z') = \frac{1}{2\kappa_{0}} \left[
e^{-\kappa_{0}|z-z'|} - \frac{c_{0}-\kappa_{0}}{c_{0}+\kappa_{0}}
e^{-\kappa_{0}(z+z')} \right],\label{11} \ee
 where
$\kappa_{0}=\sqrt{p^{2}+\mu_{0}^{2}} $ with $p$ being the value of
the parallel momentum ${\bf p}$ associated with the $d-1$
translationally invariant directions in the system.

There are two special cases: a) when two ends of the polymer are
attached to the wall (in such a case we have to deal with a
calculation of two point correlation function
$G^{(0,2)}(r,z=0;r',z'=0)$), and b) when one end of the polymer is
unrestricted in the bulk and the other one is attached to the wall
$(G^{(1,1)}(x;r',z'=0)$). In order to obtain the universal surface
critical exponents, characterizing the adsorption on the wall of
long-flexible polymer chains inserted into the medium with
long-range-correlated quenched disorder, it is sufficient to
consider the correlation function of two surface fields
$G^{(0,2)}(r,z=0;r',z'=0)$ (see \cite{DSh98}). The universal
surface critical exponents for such systems depend on the
dimensionality of space $d$, the number of order parameter
components $m (m\to 0)$ and the range of the disorder
correlations, i.e. on  $a$.

In the theory of semi-infinite systems the bulk field $\phi({\bf
x})$ and the surface field $\phi_{s}({\bf r})$ should be
reparameterized by different uv-finite renormalization factors
$Z_{\phi}(u,v,w)$ and $Z_{1}(u,v,w)$ \cite{DD81,DSh98},

$$ \phi(x) =
Z_{\phi}^{1/2}\phi_{R}(x)\quad\quad\quad {\rm and} \quad\quad\quad
 \phi_{s}(r)=Z_{\phi}^{1/2}Z_{1}^{1/2}\phi_{s,R}(r). $$
 Introducing the additional surface operator insertions $\frac{1}{2}\phi_{s}^{2}({\bf
 R}_{l})$ requires additional specific renormalization factor $Z_{\phi_{s}^2}$
 $$ \phi_{s}^2=[Z_{\phi_{s}^2}]^{-1}\phi_{s,R}^{2}. $$
The renormalized correlation function involving $N$ bulk, $M$
surface fields and $L_{1}$ surface operators
$\frac{1}{2}\phi_{s}^2({\bf R}_{l})$ can be written as
\be
 G_{R}^{(N,M,L_{1})} ({\bf{p}} ;
\mu,u,v,w,c)=Z_{\phi}^{-(N+M)/2} Z_{1}^{-M/2}
Z_{\phi_{s}^2}^{L_{1}} G^{(N,M,L_{1})} ({\bf{p}} ;
\mu_{0},u_{0},v_{0},w_{0},c_{0}).\label{12} \ee

It should be mentioned that the typical bulk short-distance
singularities, which are present in the correlation function
$G^{(0,2)}$, can be subtracted after performing the mass
renormalization. For distinguished parallel and perpendicular
directions we obtain:

\be \mu_{0}^{2}=\mu^{2}-t_{1}^{(0)} I_{1}(\mu^{2})+ t_{2}^{(0)}
I_{2}(\mu^{2}),\label{14} \ee
where
\be
t_{1}^{(0)}=\frac{1}{3}(v_{0}-u_{0}-\frac{w_{0}\mu^{a-d}}{cos[\pi(a-d)/2]}),
\quad\quad\quad
t_{2}^{(0)}=\frac{w_{0}}{3\sqrt{\pi}}\frac{\Gamma(\frac{d-a-1}{2})}{\Gamma(\frac{d-a}{2})},\label{14a}
\ee
and
\be
I_{1}(\mu^2)=\frac{1}{(2\pi)^{d-1}}\int\frac{d^{d-1}q}{2\kappa_{q}}\label{I1}
\ee
 with $\kappa_{q}=\sqrt{q^2+\mu^2}$ and
$$I_{2}(\mu^2)=\frac{1}{(2\pi)^{d-1}}\int d^{d-1}q \frac{\mid q\mid^{a-d+1}
{}_2F_{1}[\frac{1}{2},1,\frac{3+a-d}{2},\frac{q^2}{\kappa_{q}^2}]}{2
\kappa_{q}^2}.$$
According to the above mentioned notations, we have only two
coupling constants, $V_{0}=v_{0}-u_{0}$ and $w_{0}$ in the
effective Hamiltonian (we keep notation $v_{0}$ for $V_{0}$).

The renormalized coupling constants $v$, $w$
are fixed via the standard normalization conditions of the
infinite-volume theory \cite{Holovatch01}
\bea
\mu^{4-d}v&=&\Gamma_{b,R,v}^{(4)}(\{q\};\mu^2,v,w)|_{q=0},\nonumber\\
\mu^{4-a}w&=&\Gamma_{b,R,w}^{(4)}(\{q\};\mu^2,v,w)|_{q=0},\label{14b}
\eea
where $\Gamma_{b,R,v}^{(4)}$ and $\Gamma_{b,R,w}^{(4)}$ are the
$v$ and $w$ term symmetry contributions to the four-point vertex
function. To the present accuracy of calculation at one-loop
order, the vertex renormalization gives: $v=v_{0}\mu^{d-4}$ and
$w=w_{0}\mu^{a-4}$.

 In order to remove the
short-distance singularities of the correlation function
$G^{(0,2)}$, located in the vicinity of the surface, the
surface-enhancement shift $c_{0}=c+\delta c$ is required. In
accordance with this, a new normalization condition should be
introduced for the surface-enhancement shift $\delta c$ and the
surface renormalization factor $Z_{1}$.
By analogy
with magnetic systems \cite{DSh98,Sh97,UH02}, the renormalized
surface two-point correlation function in our case is normalized
in such a manner \cite{DSh98} that at zero external momentum it
should coincide with the lowest order perturbation expansion of
the surface susceptibility $\chi_{\parallel}(p)=G^{(0,2)}(p)$
 \be
G^{(0,2)} (p;\mu_{0},v_{0},w_{0},c_{0}) =
\frac{1}{c_{0}+\sqrt{p^{2} + \mu_{0}^{2}}} +
O(v_{0},w_{0}).\label{15} \ee
Thus, we obtain the necessary surface normalization condition,
\be G_{R}^{(0,2)}(0;\mu,v,w,c) = \frac{1}{\mu+c},\label{16} \ee
and for the first derivative with respect to $p^2$ we have
\be \left.\frac{\partial G_{R}^{(0,2)} (p;\mu,v,w,c)}{\partial
p^{2}} \right|_{p=0} = - \frac{1}{2\mu(\mu+c)^{2}}. \label{17} \ee
 Eq. (\ref{16}) defines the required surface-enhancement
shift $\delta c$ and shows that the surface susceptibility
diverges at $\mu=c=0$. This point corresponds to the multicritical
point $(\mu_{0c}^{2},c_{0}^{a})$, at which the adsorption
threshold takes place (it corresponds to the special transition).

From the normalization condition of Eq. (\ref{17}) and the
expression for the renormalized correlation function of Eq.
(\ref{12}), we can find the renormalization factor $Z_{\parallel}
= Z_{1} Z_{\phi}$ from the relation

\be Z_{\parallel}^{-1} = \left. 2\mu \frac{\partial}{\partial
p^{2}}[G^{(0,2)} (p)]^{-1}\right|_ {p^2=0} = \lim_{p\to
0}{\mu\over p}{\partial\over\partial p}
[G^{(0,2)}(p)]^{-1}.\label{18} \ee
The normalization condition for the correlation function
$G^{(0,2,1)}$, with the insertion of the surface operator
$\frac{1}{2}\varphi^2_{s}$,
\begin{equation}
\left.G_{R}^{(0,2,1)}({\bf{p}};\mu,u,v,c)\right|_{{\bf{p}}=0}=\frac{1}{(\mu+c)^2}\label{18a}
\end{equation}
gives a possibility of obtaining the renormalization factor
$Z_{\phi_{s}^2}$ from
\begin{equation}
[Z_{\phi_{s}^2}]^{-1}=\left. Z_{\parallel}\frac{\partial
[G^{(0,2)}(0;\mu_{0},u_{0},v_{0},c_{0})]^{-1}}{\partial
c_{0}}\right|_{c_{0}=c_{0}(c,\mu,u,v)}.\label{18b}
\end{equation}
Eq.(\ref{18a}) follows from the fact that the bare correlation
function $G^{(0,2,1)}(0;\mu_{0},u_{0},v_{0},c_{0})$ may be written
as a derivative $-\frac{\partial}{\partial
c_{0}}G^{(0,2)}(0;\mu_{0},u_{0},v_{0},c_{0})$.

\subsection{The Callan-Symanzik equations}

Asymptotically close to the critical point the renormalized
correlation functions $G_{R}^{(N,M)}$ satisfy the corresponding
homogeneous Callan-Symanzik (CS) equations \cite{Sh97,DSh98}
\begin{eqnarray}
&& \left[ \mu\frac{\partial}{\partial \mu}+\beta_{v}
(v,w)\frac{\partial}{\partial v}+ \beta_{w}
(v,w)\frac{\partial}{\partial
w}+\frac{N+M}{2}\eta (v,w)\right.\nonumber\\
&& +\left.\frac{M}{2}\eta^{sp}_{1}(v,w)\right]
G^{(N,M)}_{R}(0;\mu,v,w,c)= 0,\label{19}
\end{eqnarray}

where the $\beta$-functions are
$\beta_{v}(v,w)=\left.\mu\frac{\partial}{\partial
\mu}\right|_{L\!R}
v,\quad\beta_{w}(v,w)=\left.\mu\frac{\partial}{\partial
\mu}\right|_{L\!R} w $,

the exponents $\eta$ and $\eta_{1}^{sp}$ are

\be
 \eta=\mu\!{\partial\over \partial \mu}
\left.\ln\!{Z_{\phi}}\,\right|_{L\!R},\quad\quad\quad
\eta_{1}^{sp}=\mu\!{\partial\over \partial \mu}
\left.\ln\!{Z_{1}}\,\right|_{L\!R},\label{19b}
\ee
 and where LR is the long-range fixed point.
 It should be mentioned that up to one
loop order in the $\epsilon$- and $\delta$- expansion the LR is
located in the region of irrelevant disorder $a>3$, and up to two
loop order the LR {\it stable fixed point} is found after
performing the Borel-Chisholm resummation
\cite{Blavatska01,Holovatch01}.

The simple scaling dimensional analysis of $G_{R}^{(0,2)}$ and of
the mass dependence of the $Z$ factors, allows to express the
surface correlation exponent $\eta_{\parallel}^{sp}$ as

\be \eta_{\parallel}^{sp}=\eta_{1}^{sp}+\eta. \label{20} \ee
 From Eqs. (\ref{18}),(\ref{19b}) and (\ref{20}), we obtain for the
surface correlation exponent $\eta_{\parallel}^{sp}$ the following
expression

\begin{eqnarray}
\eta_{\parallel}^{sp}&=&\mu\!{\partial\over \partial \mu}
\left.\ln\!{Z_{\parallel}}\,\right|_{L\!R}\nonumber\\
&=&\beta_v(v,w){\partial\ln Z_{\parallel}(v,w)\over \partial v}+
\left.\beta_w(v,w){\partial\ln Z_{\parallel}(v,w)\over \partial
w}\,\right|_{L\!R},\label{21}
\end{eqnarray}
where the $\beta$ functions are \cite{Holovatch01} \bea
\beta_{\bar v}({\bar v},{\bar
w})&=&-{\bar{v}}+{\bar{v}}^2-(3f_{1}(a)-f_{2}(a)){\bar{v}}
{\bar{w}}-...,\nonumber\\
\beta_{\bar w}({\bar v},{\bar w})&=&-(4-a){\bar
w}-(f_{1}(a)-f_{2}(a)){\bar w}^2+\frac{{\bar v} {\bar
w}}{2}+...\label{22} \eea
In the above equation the renormalized coupling constants $v$ and
$w$ are normalized in a standard fashion, so that
$$
{\bar v}=\frac{4}{3}v I_{1},\quad\quad\quad {\bar w}=\frac{4}{3} w I_{1},
$$
and the integral $I_{1}$ in the case of $d=3$ is equal to $1/8\pi$
and in the case of $d=4-\epsilon$ it is
$I_{1}=2^{-d}\pi^{-d/2}\Gamma (\epsilon/2)$. The coefficients
$f_{i}(a)$, expressed via the one-loop integrals
\cite{Holovatch01,Prudnikov} are given by
\bea f_{1}(a)&=&\frac{(a-2)(a-4)}{2 \sin (\pi a/2)},\nonumber\\
f_{2}(a)&=&\frac{(a-2)(a-3)(a-4)}{48 \pi \sin
(\pi(a/2-1))}.\label{23} \eea

\subsection{crossover between the adsorbed and nonadsorbed states}

As already discussed in the introduction, it is particularly
interesting to investigate the adsorption threshold and the
crossover behavior between the adsorbed and the nonadsorbed states,
where the distribution of monomers in the near-surface region changes
character.
 In order to investigate the crossover behavior from the
nonadsorbed region, $c>c_{0}^{a}$, to the adsorbed one,
$c<c_{0}^{a}$, let us consider a small deviation $\Delta
c_{0}=c_{0}-c_{0}^{a}$ from the multicritical point. The power
series expansion of the bare correlation functions
$G^{(N,M)}({\bf{p}};\mu_{0},v_{0},w_{0},c_{0})$ in terms of  this
small deviation $\Delta c_{0}$ has the form
\begin{equation}
G^{(N,M)}({\bf{p}};\mu_{0},v_{0},w_{0},c_{0})=\sum_{L_{1}=0}^{\infty}
\frac{(\Delta c_{0})^{L_{1}}}{L_{1} !}
G^{(N,M,L_{1})}({\bf{p}};\mu_{0},v_{0},w_{0},c_{0}^{a}).\label{24}
\end{equation}
Taking into account Eq.(\ref{12}), we can rewrite the right-hand
side of Eq.(\ref{24}) in terms of the renormalized correlation
functions and  renormalized variable
 $ \Delta c=[Z_{\phi^{2}_{s}}(v,w)]^{-1} \Delta c_{0} $. In this
 way we obtain
\begin{eqnarray}
&& Z_{\phi}^{-(N+M)/2}
(Z_{1})^{-M/2}G^{(N,M)}({\bf{p}};\mu_{0},v_{0},w_{0},c_{0})\nonumber\\
&& = \sum_{L_{1}=0}^{\infty}\frac{(\Delta c)^{L_{1}}}{L_{1} !}
G^{(N,M,L_{1})}_{R}({\bf{p}};\mu,v,w).\label{25} \end{eqnarray}
The above equation determines in a straightforward fashion the
corresponding renormalized correlation functions in the vicinity
of the multicritical point $(\mu^2_{0c},c_{0}^{a})$,
\begin{equation}
G^{(N,M)}_{R}({\bf{p}};\mu,v,w,\Delta c)=Z_{\phi}^{-(N+M)/2}
(Z_{1})^{-M/2}G^{(N,M)}({\bf{p}};\mu_{0},v_{0},w_{0},c_{0}).\label{26}
\end{equation}
These correlation functions depend on the dimensionless variable $
\bar{c}=\Delta c/\mu.$ The correlation functions
$G^{(N,M)}_{R}({\bf{p}};\mu,v,w,\Delta c)$ satisfy the CS equations
(\ref{19})(see also Ref.
\cite{Sh97,DSh98}) with the additional surface related term
$-[1+\eta_{\bar{c}}(v,w)] \bar{c}\frac{\partial}{\partial
\bar{c}}$, where
\begin{equation}
\eta_{\bar{c}}(v,w)=\left. \mu\frac{\partial}{\partial
\mu}\right|_{LR} ln
Z_{\phi_{s}^2}(v,w)=\left.\beta_{v}(v,w)\frac{\partial ln
Z_{\phi_{s}^2}(v,w)}{\partial v}+ \beta_{w}(v,w)\frac{\partial ln
Z_{\phi_{s}^2}(v,w)}{\partial w}\right|_{LR}\label{27}
\end{equation}
should be calculated at the LR stable fixed point.

The asymptotic scaling critical behavior of the correlation
functions can be obtained through a detailed analysis of the CS
equations, as was proposed in Ref. \cite{Z89,BB81} and employed in
the case of the semi-infinite systems in \cite{CR97,DSh98}. Taking
into account the scaling form of the renormalization factor
$Z_{\phi_{s}^2}$ of Eq. (\ref{18b}) and the relation $\mu\sim
\tau^{\nu}$, where $\tau=(T-T_{c})/T_{c}$ is the reduced critical
temperature in magnetic systems, we obtain for $\Delta c$ and for
the scaling variable $\bar{c}$ the following asymptotic forms
\begin{equation}
\Delta c\sim \mu^{-\eta_{\bar{c}}(v^{*},w^{*})} \Delta
c_{0},\quad\quad \Delta c\sim \tau^{-\nu
\eta_{\bar{c}}(v^{*},w^{*})} \Delta c_{0}\label{28} \end{equation}
and
\begin{equation}
\bar{c}\sim \mu^{-(1+\eta_{\bar{c}}(v^{*},w^{*}))} \Delta
c_{0},\quad\quad\quad \bar{c}\sim \tau^{-\Phi} \Delta
c_{0},\label{29}
\end{equation}
where
\begin{equation}
\Phi=\nu (1+\eta_{\bar{c}}(v^{*},w^{*})) \label{30}
\end{equation}
is the surface crossover critical exponent. Eq. (\ref{29})
explains the physical meaning of the surface crossover exponent as
a value which characterizes the measure of deviation from the
multicritical point.

Taking into account the above mentioned results, we obtain from
the CS equation the following asymptotic scaling form of the
surface correlation function $G^{(0,2)}$,
\begin{eqnarray}
&& G^{(0,2)}(p;\mu_{0},v_{0},w_{0},c_{0})\sim
\mu^{-\frac{\gamma_{11}^{sp}}{\nu}}
G^{(0,2)}_{R}(\frac{p}{\mu};1,v^{*},w^{*},\mu^{-\Phi / \nu}\Delta
c_{0})\nonumber\\ && \sim
\tau^{-\gamma_{11}^{sp}}G(p\tau^{-\nu};1,\tau^{-\Phi}\Delta
c_{0}), \label{29b}
\end{eqnarray}

where $ \gamma_{11}^{sp}=\nu (1-\eta_{\parallel}), $ and $
\eta_{\parallel}^{sp}=\eta_{1}^{sp}+\eta $ are the surface
exponents at the multicritical point. The knowledge of
$\eta_{\bar{c}}$ gives access to the calculation of the critical
exponents $\alpha_{1}$ and $\alpha_{\parallel}$ of the layer and
local specific heats via the usual scaling relations \cite{D86}
\begin{equation}
\alpha_{1}=\alpha+\nu-1+\Phi=1-\nu (d-2-\eta_{\bar{c}}),\quad\quad
\alpha_{\parallel}=\alpha+\nu-2+2\Phi = -\nu
[d-3-2\eta_{\bar{c}}].\label{31}
\end{equation}

\section{The perturbation expansion for the surface critical exponents}
Applying the field-theoretical renormalization group (RG) approach
we perform calculations in a double expansion in $\epsilon=4-d$
and in $\delta=4-a$ up to the linear approximation, as was
proposed by Weinrib and Halperin \cite{WH} for infinite systems.
Thus, after performing the integration of the corresponding
Feynman integrals in the renormalized two point correlation
function $G^{(0,2)}$, we obtain at the first order of the
perturbation theory the following result for the renormalization
factor $Z_{\parallel}$
\be Z_{\parallel}=1+\frac{{\bar v}}{4 (1+\epsilon)}-\frac{{\bar
w}}{2 \cos(\frac{\pi}{2}(\delta-\epsilon))}g(\epsilon,\delta), \ee
where \bea
g(\epsilon,\delta)&=&(1-\frac{\delta}{\epsilon})\frac{1}{1+\delta}+\epsilon/2-\delta-\frac{\epsilon(\frac{1}{\delta}+\frac{\gamma_{E}}{2}+\psi_{E}
(\frac{1}{2}))}{2\Gamma(\delta-\epsilon)}
\times\nonumber\\
&&\left({}_{p}F_{q}[\{\frac{1}{2},1,2-\frac{\delta}{2}\},\{2+\frac{\delta}{2},
\frac{3+\epsilon-\delta}{2}\},1]+\right.\nonumber\\
&+&\left.{}_{p}F_{q}[\{1,\frac{3}{2},2-\frac{\delta}{2}\},\{2+\frac{\delta}{2},
\frac{3+\epsilon-\delta}{2}\},1]\right).\label{eps1} \eea
Combining the renormalization factor $Z_{\parallel}$ together with the
corresponding $\beta$-functions derived in Ref.
\cite{Holovatch01,Blavatska01}, we obtain for the surface critical
exponent $\eta_{\parallel}^{sp}$ the result
\be \eta_{\parallel}^{sp}=-\frac{{\bar
v}}{4}\frac{\epsilon}{(1+\epsilon)}+\frac{{\bar w} \delta}{2
\cos(\frac{\pi}{2}(\delta-\epsilon))}g(\epsilon,\delta)
.\label{eps2} \ee
The above mentioned surface critical exponent
$\eta_{\parallel}^{sp}$ in the case of $\epsilon,\delta$ -
expansion can be calculated formally at the corresponding fixed
point $v^{*}=\frac{2\delta^2}{(\epsilon-\delta)}$,
$w^{*}=-\frac{\delta(\epsilon-2\delta)}{(\epsilon-\delta)}$
obtained in the first order of $\epsilon,\delta$ - expansion in
\cite{Blavatska01}.

 In the special case of three spatial dimensions $d=3$ and for
arbitrary $a$ the renormalization factor $Z_{\parallel}$ at the
one-loop order is given by
 \begin{equation} Z_{\parallel}= 1 + \frac{{\bar
v}}{8}+\frac{{\bar w}g(a)}{8\sin (\pi a/2)},\label{Z11}
\end{equation}
where we have introduced the function $g(a)$ by
\bea
g(a)&=&2^{a-3}\left(\frac{\Gamma(\frac{5-a}{2})}{\Gamma(\frac{7-a}{2})}
+(3-a)\frac{\Gamma(\frac{3-a}{2})}{\Gamma(\frac{5-a}{2})}\right)\nonumber\\
&-&\frac{(3-a)}{(4-a)}
\left(\frac{\Gamma(\frac{5-a}{2})}{\Gamma(\frac{7-a}{2})}
+\frac{\Gamma(\frac{3-a}{2})}
{\Gamma(\frac{5-a}{2})}\right).\label{ga} \eea
From the Eq.(\ref{ga}) it is easy to see that in the case of
short-range correlated (or uncorrelated) disorder, i.e. for
 $a=d=3$, the above mentioned function  reduces to $g(a)=1$, and both $\bar v$
 and $\bar w$ terms are of the same symmetry. It confirms that a
 short-range correlated (or uncorrelated) disorder is irrelevant
 for SAWs.

 Combining the renormalization factor $Z_{\parallel}$
together with the one-loop pieces of the $\beta$-functions, according
to Eq.(\ref{21}), we finally obtain the following expression for
the surface critical exponent $\eta_{\parallel}^{sp}$,
\be \eta_{\parallel}^{sp}=-\frac{\bar v}{8} - \frac{\bar w}{8}
\frac{(4-a)g(a)}{\sin (\pi a/2)}.\label{etap} \ee
Similarly, for the renormalization factor $Z_{\phi_{s}^2}$ we
obtain at  the one-loop order

\be Z_{\phi_{s}^2}=1+\frac{\bar v}{2}(ln 2-\frac{1}{4})+
\frac{\bar w}{2\sin(\pi a/2)}(h(a)-\frac{g(a)}{4}),\label{32} \ee
where $h(a)$ is a combination of the Appell hypergeometric
functions of two variables $F_{1}[a,b_{1},b_{2},c,x,y]$,
\bea h(a)&=&
2^{a-4}F_{1}[1,1,\frac{5-a}{2},2,-1,-1/2]+\nonumber\\
&+&\frac{2^{(a-1)/2}}{5-a}F_{1}[\frac{5-a}{2},1,\frac{5-a}{2},
\frac{7-a}{2},-1,-2]
-\frac{4(3-a)}{(7-a)(6-a)}.
\label{33}
\eea
Finally, for the exponent $\eta_{\bar{c}}$ we obtain
\be \eta_{\bar{c}}=-\frac{\bar{v}}{2}(ln
2-\frac{1}{4})-\frac{\bar{w}(4-a)} {2 \sin(\pi
a/2)}(h(a)-\frac{g(a)}{4}).\label{34} \ee
In the case of the  short-range correlated (or uncorrelated) disorder
for the function $h(a)$ at $a=d=3$ we obtain $ln 2$.

 The above values of the surface critical exponents $\eta_{\parallel}^{sp}$
 and $\eta_{\bar{c}}$ should be calculated
at the long-range (LR) stable fixed point obtained in Ref.
\cite{Holovatch01,Prudnikov}
 for different fixed values of the correlation parameter,
$2<a\leq 3$. The  other surface critical exponents can
be calculated on the basis of the surface scaling relations (see the
Appendix) and one-loop series for the bulk critical exponents obtained
in Ref.\cite{Blavatska01,Holovatch01},

\bea
\nu^{-1}&=&2-\frac{\bar v}{4}+\frac{f_{1}(a)-f_{2}(a)}{2}{\bar
w}+...,\nonumber\\
\eta&=&\frac{1}{2}f_{2}(a){\bar w}+...\label{nueta} \eea

The results of our calculation of $[1/0]$ and $[0/1]$
Pad\'e-approximants of the series of the surface critical exponents at
the adsorption threshold, and a group of
critical exponents connected with the crossover exponent $\Phi$
are presented in Table 1 and Table 2, respectively.

In the case $a=d=3$, which corresponds to
random uncorrelated  point-like disorder (or short-range-correlated
disorder) the obtained one-loop results for the surface critical
exponents coincide with the results for the pure model (see
\cite{DSh98,Sh97} Pad\'e approximants [1/0]and [0/1]), as they
should. For $2.3\leq a < 3$ the surface critical exponents,
belonging to the new universality class associated with the LR
fixed point of the RG equations, depend on $a$, similarly as the
bulk exponents. This fact indicates that all the characteristics
of the process of adsorption on a clean wall, described in the
introduction, depend on the range of the correlations between the
defects in the bulk.

\section{discussion of the results}

Let us first discuss the effect of the long-range correlated disorder
on the distribution of monomers at and above the adsorption threshold
($c\ge 0$). From Eq.(\ref{xi>0}) and the fact that $\nu(a)$ is a
decreasing function \cite{Blavatska01,Holovatch01}, it follows that
the polymer with one end attached to the surface swells as in the bulk
when $a$ decreases. The behavior of the average end-to-end distance at
and above the threshold  is independent of the surface critical
exponents (see Eq. (\ref{xi>0})), and the relevant exponent $\nu(a)$ has been
obtained at two loop order
\cite{Blavatska01,Holovatch01}.

 Right at the threshold the average number of polymer ends and the
 average number of monomers depend on the distance from the
 surface. The distribution of monomers at different distances from the
 surface at the adsorption threshold, as well as the crossover
 behavior to the adsorbed state, can only be determined with the help
 of the surface critical exponents, as discussed in some detail in the
 introduction.  The latter exponents have been calculated in this work
 up to one loop order and we can describe the effect of the long-range
 correlated disorder on the basis of our results. Our one-loop results
 show that for decreasing $a$ the exponent $\gamma_1(a)$ strongly
 increases, whereas $\Phi(a)$ decreases. We should point out that we
 cannot exclude the possibility that the dependence of the surface
 critical exponents on $a$ is different beyond the one-loop
 approximation.

\begin{figure}[htb]
\begin{center}
\includegraphics[scale=0.8]{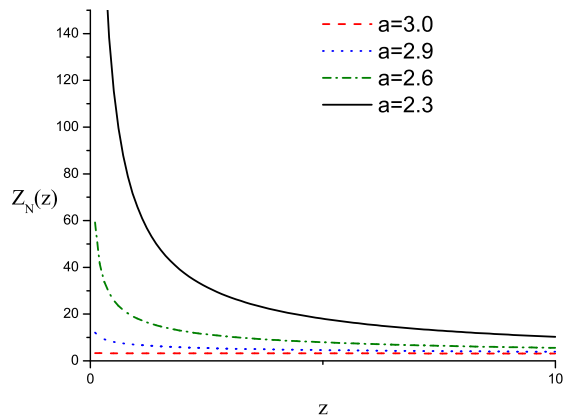}
\end{center}
\caption{The  partition function $Z_{N}(z)$  just at  the adsorption
 thereshold $c=0$ and for $N=100$, as a function of $z/l$ for $1\ll
 z/l\ll N^{\nu}$ and for different values of $a$.  $Z_{N}(z)$ is
 dimensionless and $l$ is the microscopic length scale.
 }\label{fig1}
\end{figure}

\begin{figure}[htb]
\begin{center}
\includegraphics[scale=0.8]{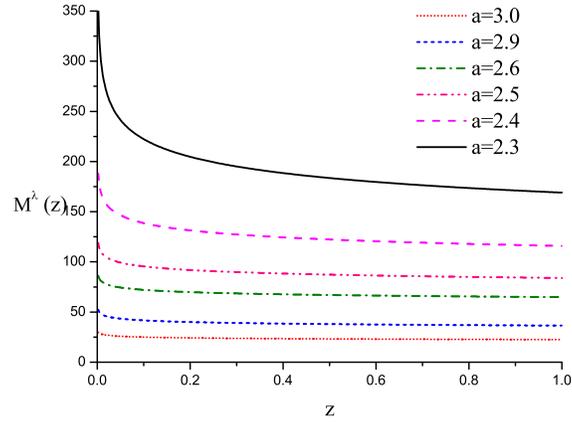}
\end{center}
\caption{The  density of monomers $M^{\lambda}(z)$ in the layer at the
 distance $z$ from the surface to which one end of the chain is
 attached for $1\ll z/l \ll N^{\nu}$ just at the threshold $c=0$ and for
 $N=100$ for different values of
 $a$. $M^{\lambda}(z)$ is in arbitrary units and $z/l$ is
 dimensionless.}
\label{fig2}
\end{figure}

Let us describe the effect of the long-range correlated disorder at
the adsorption threshold and at the crossover to the adsorbed state
assuming that the qualitative trends are properly captured by the
one-loop results.  For small distances ($l\ll z\ll \xi_R$) and for
$c=0$ the partition function $Z_N(z)$ and the number of monomers in
the layer at $z$ are shown in Figs. \ref{fig1} and \ref{fig2}
respectively for several values of $a$. From these plots (see also
Eqs. (\ref{ZN}) and (\ref{MN}) and Tables I, II and Table in
Ref.\cite{Blavatska01}) we can see that the number of free ends and
the number of monomers in the near-surface region ($l\ll z\ll \xi_R$)
both increase for deceasing $a$. Moreover, the dependence on $z$ is
the stronger, the larger the range of correlations between the
defects. This result suggests that the larger is the range of
correlations between the defects, the more efficient is the trapping
of the chain between the attractive surface and the region occupied by
the defects. For smaller $a$ more steps are necessary for the chain to
go around the defected region, which is larger for smaller $a$. From
the fact that $\Phi$ decreases for decreasing $a$ it follows that the
fraction of the monomers adsorbed at the surface also decreases (see
Eq.(\ref{N1/N}) and Table II). This result is somewhat surprising,
since it shows a trend just opposite to the one for the fraction of
monomers contained in the layer at the distance $z$ from the wall for
$l\ll z\ll
\xi_R$. A possible explanation might be that once the polymer leaves
the surface, the large disordered patches in the bulk make it
difficult for the polymer to return back. As a result, the fraction of
the monomers near the wall can be higher than right at the
wall. Recall that also in the pure systems for $c=0$ there is a discrepancy
between the behavior of the fraction of monomers right at the surface,
$N_1/N\sim N^{\Phi-1}$, and the fraction of monomers at the distance
from the surface of the order of the effective monomer dimension $l$,
$M^{\lambda}_N(l)/N\sim N^{\Phi-1+\gamma_1-1}$.  Since $\gamma_1>1$
beyond MF, the concentration of monomers close to the surface is
larger than just at the boundary, and $N_1/M^{\lambda}_N(l)\sim
N^{1-\gamma_1}\to 0$ for $N\to\infty$ and $c=0$. Our one-loop results
indicate that this effect is enhanced in the presence of long-range
correlated disorder.

\begin{figure}[htb]
\begin{center}
\includegraphics[scale=0.9]{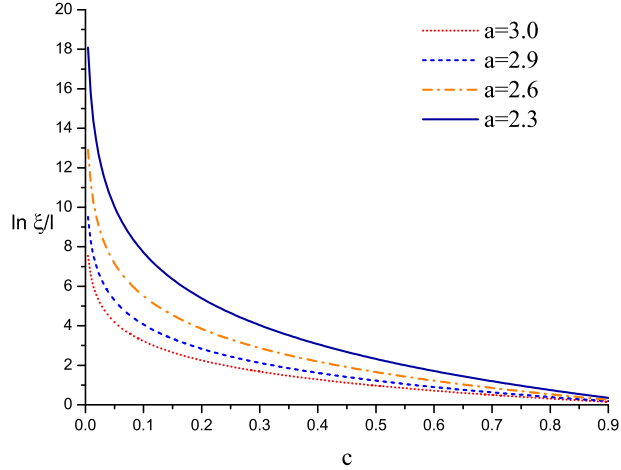}
\end{center}
\caption{The dependence of the thickness of the adsorbed layer
$\xi/l$ on $|c|$, for $c<0$, where $c\propto (T-T_a)/T_a$ is the
reduced temperature distance from the threshold, for different
values of $a$. Both quantities are dimensionless. }
 \label{fig3}
\end{figure}

Let us now consider the crossover to the adsorbed state. Assume
that one end of the infinite polymer chain is attached to the surface, and the
temperature is decreased below the threshold, i.e. $-1<c<0$.
 The thickness of the adsorbed layer (Eq.(\ref{xi}))
starts to decrease from infinity when $|c|$ increases from zero
(see Fig.\ref{fig3}).
 From Eq.(\ref{xic}) and Table II we see that
for a given $c$ such that $|c|< 1$, the thickness of the
near-surface, polymer layer increases when the range of
correlations increases ($a$ decreases). Moreover, the dependence
on $a$ of the exponent $\nu(a)/\Phi(a)$ is stronger than in the
case of $\nu(a)$.  Hence, the dependence of the thickness of the
adsorbed layer $\xi\sim |c|^{-\nu(a)/\Phi(a)} $ on $a$ is stronger
than the corresponding dependence of the mean end-to-end distance
$\xi_R\sim N^{\nu(a)}$ in the bulk. It indicates that the effect
of the long-range disorder on the adsorbed layer just below the
threshold is stronger than the corresponding effect on the polymer
coil in the bulk. The fraction of monomers, on the other hand,
decreases for a fixed temperature for decreasing $a$, because for
decreasing $a$, $\Phi(a)$ decreases (see Eq. (\ref{N1/N}) and
Table II).

\begin{figure}[htb]
\begin{center}
\includegraphics[scale=0.8]{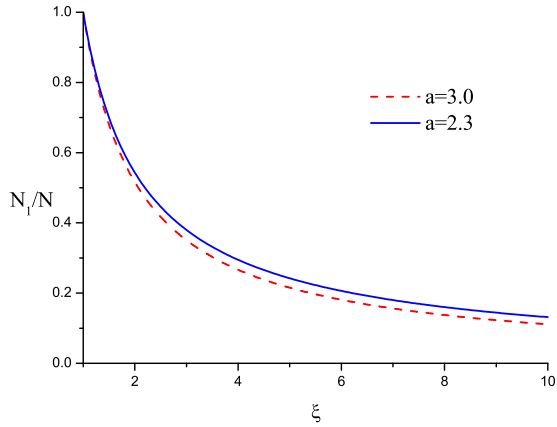}
\end{center}
\caption{The dependence of the fraction of  monomers at the
surface $N_1/N$ on the thickness of the adsorbed layer $\xi/l$ for
$c<0$ (i.e. below the threshold) and for two values of $a$,
$a=3.0$ and $a=2.3$. Both quantities are dimensionless.}
\label{fig4}
\end{figure}

In Fig.\ref{fig4} the dependence of the fraction of monomers on
the thickness of the adsorbed layer just below the threshold is
shown for two values of $a$. Note that for a given value of the
fraction of monomers at the surface the thickness of the adsorbed
layer increases for decreasing $a$. Let us compare two systems,
characterized by different ranges of disorder-correlations,
$a_1<a_2$. Each system contains a chain with $N$ monomers. Assume
finally that the same number of monomers $N_1$ is adsorbed at the
surface in each system (of course temperatures in these systems
are different, see (\ref{N1/N}) and Table II).
 The number of monomers
inside the solution, $N-N_1$ is the same in the two considered
cases. Since $a_1<a_2$, larger correlated patches have to be avoided
by the chain in the first system.  By analogy with the polymer
properties in the bulk
\cite{Blavatska01} it is natural to expect that the same number of monomers
contained in the solution, $N-N_1$, must effectively occupy larger
space in the case where larger correlated patches have to be
surrounded by the chain. Our results are thus consistent with the
former results for the bulk systems \cite{Blavatska01}.

We conclude that the long-range correlated disorder has a
significant effect on the adsorption of polymers on the surface at
and near the adsorption threshold. When one end of the polymer is
attached to the surface, the perpendicular part of the average
distance to the other end increases for increasing range of
correlations between the disorder. At the same time, the fraction
of monomers at the surface decreases for fixed temperature both at
and below the threshold. Moreover, just at the threshold the
monomer concentration near the wall (i.e. for $l\ll z\ll \xi_R$)
increases. We should stress that the above conclusions follow from
the results obtained in the one-loop approximation, and the
possibility that the real dependence of the surface critical
exponents on $a$ is different cannot be excluded. Two-loop
calculations and/or computer simulations, going beyond the scope
of this work, might help to draw definite conclusions.

\begin{table}[htb]
\caption{\label{tab:tab1}Surface critical exponents of the
long-flexible polymer at the special transition $c=c_a$ for $d=3$
up to one-loop order calculated at the pure (the case $a=3$ with
$(v^{*}=1.632,w^{*}=0)$) and LR stable fixed point for different
fixed values of the correlation parameter, $2<a< 3$.}
\begin{center}
\begin{tabular}{rrrrrrrrrrr}
\hline $  a  $~&~$  v^{*} $~&~$  w^{*}  $~&~$  \eta_{\parallel}
$~&~$    $~&~$  \eta_{\perp}  $~&~$    $~&~$  \beta_{1} $~&~$
$~&~$ \gamma_{11} $~&~$ \gamma_{1}$ \\
\hline $     $~&~$ $~&~$ $~&~$ [1/0] $~&~$ [0/1] $~&~$
[1/0] $~&~$ [0/1] $~&~$ [1/0] $~&~$ [0/1] $~&~$ [1/0] $~&~$ [1/0] $ \\
\hline
 3.0  &  1.63  &  0.00  &  -0.204  & -0.169  &  -0.102  & -0.114 &  0.250
 & 0.250 &  0.704  &  1.255  \\

 2.9  &  4.13  &  1.47  &  -0.342  &  -0.255 & -0.171 & -0.146 &  0.247  & 0.248 &
 0.837 &  1.418  \\

 2.8  &  4.73  &  1.68  &  -0.402 & -0.287 & -0.200 & -0.167 &  0.244 &
 0.244 & 0.891 & 1.480 \\

 2.7  &  5.31  &  1.81  &  -0.468 & -0.319 &  -0.233 & -0.189 &  0.241 &
 0.241 &  0.951 & 1.550 \\

 2.6  &  5.89  &  1.87  &  -0.542 & -0.351 &  -0.270 & -0.212 &  0.238 &
 0.238 &  1.018 & 1.630 \\

 2.5  &  6.48  &  1.89  &  -0.620  & -0.383 &  -0.308 & -0.236 &  0.235 &
 0.236 &  1.090 & 1.715  \\

 2.4  &  7.10  &  1.87  &  -0.704 & -0.413 &  -0.350 & -0.259 &  0.233 &
 0.233 & 1.169 & 1.810 \\

 2.3  &  7.76  &  1.84  &  -0.793 & -0.442 &  -0.394 & -0.283 &  0.230 &
 0.231 &  1.253  &  1.911 \\
\end{tabular}
\end{center}
\end{table}
\begin{table}[htb]
\caption{\label{tab:tab2}Surface critical exponents of the
long-flexible polymer at the special transition $c=c_a$ for $d=3$
involving the RG function $\eta_{\bar{c}}$, calculated at the pure
(the case $a=3$) and LR stable fixed point for different fixed
values of the correlation parameter, $2<a< 3$.}
\begin{center}
\begin{tabular}{rrrrrrrrrrrrr}
\hline $  a  $~&~$  v^{*} $~&~$  w^{*}  $~&~$  \eta_{\bar{c}}
$~&~$   $~&~$  \alpha_{1} $~&~$   $~&~$  \alpha_{\parallel} $~&~$
$~&~$ \Phi $~&~$     $~&~$(1-\Phi)$~&~$/\nu$ \\
\hline $    $~&~$    $~&~$    $~&~$ [1/0] $~&~$ [0/1] $~&~$
[1/0] $~&~$ [0/1] $~&~$ [1/0] $~&~$ [0/1] $~&~$ [1/0] $~&~$ [0/1] $~&~$ [1/0] $~&~$ [0/1]$ \\
\hline
 3.0  &  1.63  &  0.00  &  -0.362  & -0.266  &  0.217  & 0.280 &
 -0.362 & -0.266 &  0.421  &  0.427  & 0.954 & 0.956 \\

 2.9  &  4.13  &  1.47  &  -0.607  &  -0.378 & 0.031 & 0.181 &  -0.607  & -0.378 &
 0.363 &  0.379 & 0.942 & 0.945 \\

 2.8  &  4.73  &  1.68  &  -0.710 & -0.415 & -0.045 & 0.147 &  -0.710 & -0.415 &
 0.335 & 0.358 & 0.950 & 0.953 \\

 2.7  &  5.31  &  1.81  &  -0.822 & -0.451 &  -0.128 & 0.114 &  -0.822 & -0.451 &
 0.306 & 0.337 & 0.955 & 0.957 \\

 2.6  &  5.89  &  1.87  &  -0.943 & -0.485 &  -0.219 & 0.082 &  -0.943 & -0.485 &
 0.276 & 0.317 & 0.954 & 0.956 \\

 2.5  &  6.48  &  1.89  &  -1.066  & -0.516 &  -0.314 & 0.051 &  -1.066 & -0.516 &
 0.247 & 0.298 & 0.944 & 0.947 \\

 2.4  &  7.10  &  1.87  &  -1.192 & -0.544 &  -0.413 & 0.023 &  -1.192 & -0.544 &
 0.222 & 0.282 & 0.922 & 0.928 \\

 2.3  &  7.76  &  1.84  &  -1.309 & -0.567 &  -0.511 & -0.003 &  -1.309 & -0.567 &
 0.203  &  0.271 & 0.881 & 0.894 \\
\end{tabular}
\end{center}
\end{table}

\renewcommand{\theequation}{A2.\arabic{equation}}
\section*{Appendix }
\setcounter{equation}{0}
 The individual RG series expansions for
the other critical exponents can be derived through standard
surface scaling relations \cite{D86} with $d=3$
\begin{eqnarray}
&& \eta_{\perp} = \frac{\eta +
\eta_{\parallel}}{2}, \nonumber\\
&& \beta_{1} = \frac{\nu}{2}
(d-2+\eta_{\parallel}), \nonumber\\
&& \gamma_{11}=\nu(1-\eta_{\parallel}), \nonumber\\
&& \gamma_{1}= \nu(2-\eta_{\perp}), \label{sc}\\
&& \Delta_{1}= \frac{\nu}{2} (d-\eta_{\parallel}), \nonumber\\
&& \delta_{1} = \frac{\Delta}{\beta_{1}} =
\frac{d+2-\eta}{d-2+\eta_{\parallel}}, \nonumber\\
&& \delta_{11} = \frac{\Delta_{1}}{\beta_{1}}=
\frac{d-\eta_{\parallel}}{d-2+\eta_{\parallel}}\;.\nonumber
\end{eqnarray}

Each of these critical exponents characterizes certain properties
of the semi-infinite systems with long-range quenched disorder, in
the vicinity of the critical point. The values $\nu$, $\eta$, and
$\Delta=\nu(d+2-\eta)/2$ are the standard bulk exponents.
\section*{Acknowledgments}
 This work was supported by NATO Science Fellowships National
 Administration under Grant No. 14/B/02/PL. One of us (AC)
 acknowledges  a partial support by the Polish Ministry of Sciences
 through the Research Project No.  4T09A06622.


\begin{thebibliography}{99}
\bibitem{Blavatska01} V.Blavats'ka, C.von Feber, and Yu.Holovatch,
J.Mol.Liq. {\bf 92}, 77 (2001).

\bibitem{Holovatch01} V.Blavats'ka,C.von Feber, Yu.Holovatch, Phys.Rev.E,
{\bf 64}, 041102 (2001).

\bibitem{deGennes} P.-G. de Gennes, Scaling Concepts in Polymer Physics
(Cornell University) Press, Ithaca, NY, 1979).

\bibitem{EKB82} E.Eisenriegler, K. Kremer, and K. Binder, J. Chem. Phys.
{\bf 77}, 12 (1982).

\bibitem{eisenriegler:83} E.Eisenriegler, J.Chem.Phys. {\bf 79}, 1052 (1983)

\bibitem{Eisenriegler} E.Eisenriegler, Polymers near surfaces, World
Scientific Publishing Co.Pte.Ltd., 1993.

\bibitem{HG94} R.Hegger and p.Grassberger, J.Phys.A {\bf 27}, 4069
(1994).

\bibitem{SKG99} Y.Singh, S.Kumar, and D.Giri, J.Phys.A {\bf 32},
L407 (1999).

\bibitem{SGK01} Y.Singh, D.Giri, and S.Kumar, J.Phys.A {\bf 34},
L67 (2001).

\bibitem{RDGKS02} R.Rajesh, D.Dhar, D.Giri, S.Kumar, and Y.Singh,
Phys.Rev.E {\bf 65}, 056124 (2002).

\bibitem{ZLB90} D.Zhao and T.Lookman, K.De'Bell, Phys.Rev.A {\bf
42}, 4591 (1990).

\bibitem{DD81} H.W.Diehl and S.Dietrich, Z.Phys.B {\bf 42}, 65 (1981).

\bibitem{DD83} H.W.Diehl and S.Dietrich, Z.Phys.B {\bf 50}, 117 (1983).

\bibitem{D86} H.~W. Diehl,  in {\em Phase Transitions and Critical Phenomena}, edited by C.
  Domb and J.~L. Lebowitz (Academic Press, London, 1986), Vol.~10, pp.\
  75--267.

\bibitem{DN89}H. W. Diehl and A. N{\"u}sser, Z. Phys. B {\bf 79}, 69
(1990).

\bibitem{Cloizeaux} J.des Cloizeaux and G.Jannink, Polymers in Solution
(Clarendon Press, Oxford, 1990); L.Sch\"afer L, Universal
Properties of Polymer Solutions as Explained by the
Renormalization Group (Springer, Berlin, 1999).

\bibitem{G76} P.G.de Gennes, J.Phys. (Paris) {\bf 37}, 1445
(1976); M.Daoud and P.G. de Gennes, ibid. {\bf 38}, 85 (1977).

\bibitem{Barber} M.N.Barber, A.S.Guttman, K.M.Middlemiss,
G.M.Torrie, and S.G.Whittington, J.Phys.A {\bf 11}, 1833 (1978).

\bibitem{DSh98}H.~W. Diehl and M. Shpot, Nucl. Phys. B {\bf 528},  595  (1998).

\bibitem{Kim} Y.Kim, J.Phys.C {\bf 16}, 1345 (1983).

\bibitem{Harris} A. B. Harris, Z. Phys. B: Condens.Matter {\bf 49}, 347
(1983).

\bibitem{Harris74}
A.~B. Harris, Journ. Phys. C {\bf 7},  1671  (1974).

\bibitem{Guida98} R.Guida and J.Zinn Justin J.Phys.A {\bf 31}, 8104
(1998).

\bibitem{Sh97}
M. Shpot, Cond. Mat. Phys. N 10, 143  (1997).

\bibitem{UH02} Z.Usatenko, Chin-Kun Hu, Phys.Rev.E {\bf 65}, 056102 (2001).

\bibitem{Prudnikov}V.V.Prudnikov, P.V.Prudnikov, and A.A.Fedorenko,
J.Phys.A {\bf 32}, L399 (1999); V.V.Prudnikov, P.V.Prudnikov, and
A.A.Fedorenko, ibid. {\bf 32},8587 (1999); V.V.Prudnikov,
P.V.Prudnikov, and A.A.Fedorenko, Phys.Rev.B {\bf 62}, 8777
(2000).

\bibitem{Z89} J.Zinn-Justin, Euclidean Field Theory and Critical Phenomena
(Oxford Univ. Press, New York, 1989).

\bibitem{BB81} C.Bagnuls and C.Bervillier, Phys.Rev.B {\bf 24}, 1226 (1981).

\bibitem{CR97} A.Ciach and U.Ritschel, Nucl.Phys.B {\bf 489}, 653 (1997).

\bibitem{WH} A. Weinrib and B. I. Halperin, Phys. Rev. B {\bf 27}, 413
(1983).

\end{thebibliography}
\end{document}